\documentclass{aastex62}

\graphicspath{{./}{figures/}}

\submitjournal{ApJ}

\usepackage{natbib}
\usepackage{graphicx,subfigure} 
\usepackage{amsmath}

\shorttitle{High resolution X-ray spectrum of quiescent Sgr~A*}
\shortauthors{Corrales et al.}

\newcommand\lsim{\mathrel{\rlap{\lower4pt\hbox{\hskip1pt$\sim$}}
        \raise1pt\hbox{$<$}}}
\newcommand\gsim{\mathrel{\rlap{\lower4pt\hbox{\hskip1pt$\sim$}}
        \raise1pt\hbox{$>$}}}

\newcommand \Msun{ {\rm M}_\odot }
\newcommand \NH{ {\rm N}_{\rm H} }

\newcommand \SgrA{Sgr~A*} 
\newcommand \SgrAEast{Sgr~A~East}
\newcommand \magnetar{SGR~J1745-29}

\newcommand \Chandra{{\sl Chandra}}

\newcommand \GCPWN{PWN~G359.945-0.045}

\begin{document}

\title{The Chandra High Resolution X-ray Spectrum of Quiescent Emission from Sgr A*}

\author[0000-0002-5466-3817]{Lia Corrales}
\affil{LSA Collegiate Fellow, University of Michigan, 
Ann Arbor, MI 48109, USA}

\author{F.~K.~Baganoff}
\affil{MIT Kavli Institute for Astrophysics and Space Research, Cambridge, MA 02139, USA}

\author{Q.~D.~Wang}
\affil{University of Massachusetts Amherst, Amherst, MA 01003, USA}

\author{M.~Nowak}
\affil{Washington University in St. Louis, St. Louis, MO 63130, USA}

\author{J.~Neilsen}
\affil{Villanova University, Villanova, PA 19085, USA}

\author{S.~Markoff}
\affil{University of Amsterdam, Anton Pannekoek Institute of Astronomy, 1090 GE Amsterdam, Nederland}

\author{D.~Haggard}
\affil{McGill University, McGill Space Institute, Montreal, QC H3A 2A7, Canada}

\author{J.~Davis}
\affil{Harvard \& Smithsonian Center for Astrophysics, Cambridge, MA 02138, USA}

\author{J.~Houck}
\affil{Harvard \& Smithsonian Center for Astrophysics, Cambridge, MA 02138, USA}

\author{D.~Principe}
\affil{MIT Kavli Institute for Astrophysics and Space Research, Cambridge, MA 02139, USA}

\begin{abstract}
In quiescence, Sgr A* is surprisingly dim, shining 100,000 times less than expected for its environment.  This problem has motivated a host of theoretical models to explain radiatively inefficient accretion flows (RIAFs).  
The {\sl Chandra} Galactic Center (GC) X-ray Visionary Program obtained approximately 3 Ms (one month) of {\sl Chandra}~HETG data, offering the only opportunity to examine the quiescent X-ray emission of \SgrA\ with high resolution spectroscopy. Utilizing custom background regions and filters for removing overlapping point sources, this work provides the first ever look at stacked HETG spectra of \SgrA. We model the background datasets with a cubic spline and fit the unbinned \SgrA\ spectra with a simple parametric model of a power law plus Gaussian lines under the effects of interstellar extinction. We detect a strong 6.7~keV iron emission line in the HEG spectra and a 3.1~keV emission line in the MEG spectra. In all cases, the line centroids and equivalent widths are consistent with those measured from low-resolution CCD spectra. An examination of the unbinned, stacked HEG$\pm 1$ spectrum reveals fine structure in the iron line complex. In addition to resolving the resonant and forbidden lines from He-like iron, there are apparent emission features arising with higher statistical significance at lower energy, potentially associated with FeXX-XXIV ions in a $\sim 1$~keV plasma arising near the Bondi radius of \SgrA. 
With this work, we release the cleaned and stacked \SgrA\ and background HETG spectra to the public as a special legacy dataset. 

\end{abstract}


\section{Introduction}
\label{sec:Intro}

The first \Chandra\ observations of the Galactic Center (GC) were essential for resolving X-ray emission from \SgrA, our Galaxy's central supermassive black hole (SMBH), from nearby X-ray binaries, pulsar wind nebulae, and stellar clusters \citep{Baganoff2001,Baganoff2003}.  
\SgrA is embedded in an environment of hot plasma, fed by the stellar winds of nearby young massive stars.  The plasma environment implies a Bondi capture rate of material $\dot{M}_b \sim 10^{-6} - 10^{-5}~\Msun$~yr$^{-1}$ \citep{Baganoff2003, Yuan2003}.  Under standard accretion assumptions, the expected bolometric luminosity is a factor of $10^5$ larger than the observed value, $L_{\rm bol} \approx 10^{36}$~erg~s$^{-1}$.  Radio polarization measurements have also shown that the density of material close to the SMBH event horizon is much smaller, implying $\dot{M} \leq 10^{-7}~\Msun$~yr$^{-1}$ \citep{Bower2003,Bower2019}.  This dilemma has inspired a series of accretion models that seek to explain both the low-luminosity and low-accretion rate of \SgrA.

Radiatively inefficient accretion flow (RIAF) is a term generally used to describe low-luminosity accretion. RIAFs are often described following a generic parameterization for mass accretion rate as a function of radius \citep[e.g.,][]{Yuan2003, Xu2006}:
\begin{equation}
	\label{eq:Mdot}
	\dot{M} \propto r^s
\end{equation}
where $\dot{M}$ is the rate of inward mass flow as a function of radius, $r$, which is the distance from the central compact object in units of the Bondi radius, and $s = 0$ corresponds to the Bondi (zero angular momentum) accretion solution. The corresponding plasma density profile is:
\begin{equation}
	\label{eq:ne}
	n_e \propto r^{-3/2 + s}
\end{equation}
and the electron temperature profile is parameterized as:
\begin{equation}
	\label{eq:Te}
	T_e \propto r^{-q}.
\end{equation}

An early model for low-luminosity accretion, and a subset of the RIAF models, is the theory of advection dominated accretion flow (ADAF). ADAFs provide an analytic solution for optically thin accretion flows in which nearly all the energy produced by viscous dissipation is transferred to ions rather than electrons, allowing for stable solutions of quasi-spherical accretion with convection and outflow with very low radiation efficiency \citep{Narayan1994,Narayan1995b,Narayan1995}. To describe the multi-wavelength spectral energy distribution (SED) of \SgrA, ADAFs require a non-thermal component of radiation, motivating jet-ADAF models \citep{Falcke1993,Yuan2002,Falcke2004}. However, the radio polarization measurements mentioned above rule out the non-outflow ADAF solution ($s=0$) \SgrA\ \citep{Agol2000,Quataert2000b}.  Models that explore outflow ADAF solutions include ADIOS \citep[adiabatic inflow-outlow,][]{BB1999} and CDAF \citep[convection dominated accretion flow][]{Quataert2000}, but neither have been shown to fully described the multi-wavelength SED of \SgrA.

\citet{Yuan2003} showed that a non-thermal population of electrons is necessary to produce the observed radio luminosity of \SgrA, requiring only a small ($\sim 1$\%) of the dissipated accretion energy going to produce a power-law tail in the electron energy distribution. In this model, a substantial fraction ($\sim 50\%$) of the viscous dissipation energy is transferred to electrons, disqualifying it from being classified as an ADAF, which assumes that the fraction is closer to 1\%. Hence, models of this flavor are typically referred to solely as RIAFs. 

When it comes to accelerating electrons in a RIAF, there are two competing scenarios. 
In the jet dominated case, the sub-mm ``bump'' in the \SgrA\ spectrum constrains the properties of the jet launching site, and a fraction of the X-ray emission comes from synchrotron self-Compton (SSC) radiation in the jet \citep{FM2000,Yuan2002}.  In the accretion flow scenario, the majority of X-ray emission comes from the quasi-spherical plasma atmosphere of \SgrA, while the radio emission comes from a non-thermal population of electrons accelerated by magnetic fields near the black hole, which would also explain the origin of flares \citep{Yuan2003,Ball2016,Roberts2017,Ma2019}.  

In a seminal work by \citet[][hereafter W13]{Wang2013}, the zeroth order CCD resolution X-ray spectrum for \SgrA\ was presented from the 3~Ms dataset obtained via the {\sl Chandra} Galactic Center X-ray Visionary Program (GCXVP). They demonstrated that the quiescent spectrum could be described by a RIAF model with a relatively flat density and temperature profile ($s \sim 1$ and $q \gtrsim 0.6$) utilizing constraints from the radio observations). These results implies a near perfect balance between outflow and inflow. They argue against jets as a source of the outflow, based on scaling relations between accretion rate and jet power, and given the low accretion rate onto \SgrA.

This work presents the high resolution HETG spectrum for quiescent \SgrA\ emission obtained from the {\sl Chandra} GCXVP 3~Ms dataset. Section~\ref{sec:Observations} describes the dataset, data reduction techniques, and parametric fits to the CCD spectrum used as benchmarks to test the quality of our HETG extraction. Section~\ref{sec:Results} describes the results of a parametric fit to the stacked HEG and MEG spectra, where we detect iron and argon lines. Section~\ref{sec:Discussion} compares our measurements for line centroid and equivalent widths to theoretical predictions made over the last 20 years, as well as a comparison to the best fit RIAF model of W13. Section~\ref{sec:Conclusions} summarizes the unexpected results and outlines avenues for future development of low-luminosity accretion flow models. The legacy dataset provided in this work is key for understanding the accretion environment of \SgrA\ and a higher signal-to-noise dataset will not be achieved for decades.

\section{Observations}
\label{sec:Observations}

All data reduction was performed using the CXC software CIAO version 4.9 with CALDB 4.7.0.  
Table~\ref{tab:hetgs_obsids} lists HETG observations from the \Chandra\ GC XVP campaign used in this project. We excluded observations with \magnetar, the magnetar $2''$ away from \SgrA\ that went into outburst in April 2013 \citet{ATEL5009}. We also limited the dataset to HETG-S observations longer than 25~ks, which were deep enough to identify enough point sources for image alignment. 
We aligned the observations by matching point sources with CIAO \texttt{celldetect} and \texttt{wcs\_merge}. We reprojected all observations onto the same sky coordinates as ObsID 13842 using \texttt{reproject\_obs}. 

We removed particle background flares by examining light curves from the ACIS-S1 (back illuminated) and ACIS-S5 (front illuminated) CCD chips. These chips are about $15''$ from the GC pointing, avoiding contamination by outbursts from \SgrA\ and most GC compact objects. The front illuminated and back illuminated chips also exhibit different magnitudes of sensitivity to particle background flares, and the background spectrum for a flare is different than that in quiescence.\footnote{CXC memo: http://cxc.harvard.edu/contrib/maxim/bg} 
We used 239.58 second bins as suggested by the CIAO background analysis threads. 
We removed time bins where the 10-14 keV count rate (due only to particle background) and the background hardness ratio (10-14 keV to 3-6 keV) varied by more than 3-sigma. This process was repeated until no additional bins were flagged. 
Examining the background hardness ratio ensured that the final dataset only includes times when the particle background spectrum is stable.

Finally, we removed \SgrA\ flares using the results of \citet{Neilsen2013}. The filtered dataset yields an average of 0.002 counts per second (2-8~keV) for \SgrA\ in quiescence. Table~\ref{tab:hetgs_obsids} lists the exposure times before and after filtering out background and \SgrA\ flares.

\begin{table} 
 \centering 
 \caption{\Chandra\ Observations used} 
 \label{tab:hetgs_obsids} 

\begin{tabular}{c c c c c} 
 \hline 
 	 & Start 	 & Raw 	 & Filt 	 & Roll \\ 
 ObsId 	 & Date 	 & Exp (ks) 	 & Exp (ks) 	 & Angle (deg) \\ 
 \hline 
13850 	 & 2012-02-06 	 & 59.3 	 & 58.0 	 & 92.2 \\ 
14392 	 & 2012-02-09 	 & 58.5 	 & 49.4 	 & 92.2 \\ 
14393 	 & 2012-02-11 	 & 41.0 	 & 36.7 	 & 92.2 \\ 
13856 	 & 2012-03-15 	 & 39.5 	 & 39.0 	 & 92.2 \\ 
13857 	 & 2012-03-17 	 & 39.0 	 & 37.7 	 & 92.2 \\ 
13847 	 & 2012-04-30 	 & 152.0 	 & 145.3 	 & 76.6 \\ 
14427 	 & 2012-05-06 	 & 79.0 	 & 72.0 	 & 76.4 \\ 
13848 	 & 2012-05-09 	 & 96.9 	 & 92.0 	 & 76.4 \\ 
13849 	 & 2012-05-11 	 & 176.4 	 & 167.7 	 & 76.4 \\ 
13846 	 & 2012-05-16 	 & 55.5 	 & 53.5 	 & 76.4 \\ 
14438 	 & 2012-05-18 	 & 25.5 	 & 24.9 	 & 76.4 \\ 
13845 	 & 2012-05-19 	 & 133.5 	 & 124.7 	 & 76.4 \\ 
14461 	 & 2012-07-12 	 & 50.3 	 & 45.4 	 & 282.3 \\ 
13853 	 & 2012-07-14 	 & 72.7 	 & 69.6 	 & 282.3 \\ 
13841 	 & 2012-07-17 	 & 44.5 	 & 42.9 	 & 282.3 \\ 
14465 	 & 2012-07-18 	 & 43.8 	 & 34.4 	 & 282.3 \\ 
14466 	 & 2012-07-20 	 & 44.5 	 & 38.4 	 & 282.3 \\ 
13842 	 & 2012-07-21 	 & 189.3 	 & 166.7 	 & 282.3 \\ 
13839 	 & 2012-07-24 	 & 173.9 	 & 145.9 	 & 282.3 \\ 
13840 	 & 2012-07-26 	 & 160.4 	 & 149.3 	 & 282.3 \\ 
14432 	 & 2012-07-30 	 & 73.3 	 & 61.4 	 & 282.3 \\ 
13838 	 & 2012-08-01 	 & 98.3 	 & 92.3 	 & 282.3 \\ 
13852 	 & 2012-08-04 	 & 154.5 	 & 133.9 	 & 282.3 \\ 
14439 	 & 2012-08-06 	 & 110.3 	 & 106.8 	 & 270.7 \\ 
14462 	 & 2012-10-06 	 & 131.6 	 & 124.2 	 & 268.7 \\ 
13851 	 & 2012-10-16 	 & 105.7 	 & 96.3 	 & 268.7 \\ 
15568 	 & 2012-10-18 	 & 35.6 	 & 33.3 	 & 268.7 \\ 
13843 	 & 2012-10-22 	 & 119.1 	 & 107.9 	 & 268.7 \\ 
15570 	 & 2012-10-25 	 & 67.8 	 & 64.0 	 & 268.7 \\ 
14468 	 & 2012-10-29 	 & 144.2 	 & 134.1 	 & 268.7 \\ 
\hline 
 \multicolumn{2}{l}{{\bf Total Exposure (Ms)}} 	 & {\bf 2.78} & {\bf 2.55} 	 &  \\ 
 \hline 
 \end{tabular} 
 \end{table}

\subsection{Low Resolution Quiescent Emission}
\label{sec:CCDspectrum}

To obtain a benchmark for comparison to the high resolution HETG spectrum, we first examined the low resolution CCD spectrum of \SgrA.  
Using the zeroth order image of \SgrA, we selected source and background regions files similar to those used by \citet[][referred to as W13 hereafter]{Wang2013}. 
We applied the ISM absorption model \texttt{tbvarabs} with ISM abundances set to those of \citet{Wilms2000}.   
We also included the effects of foreground dust scattering removing light from the extraction aperture, applying the extinction term $\exp(-\tau_{\rm sca}$) with $\tau_{\rm sca} = 0.486 (\NH/10^{22}~{\rm cm}^{-2})$ \citep{PS1995, Nowak2012, Corrales2016}. We grouped the spectrum to have a minimum of five counts per bin and used the Cash statistic for fitting, using the {\sl Interactive Spectral Interpretation System} \citep[ISIS,][]{ISISsoftware} and Heasoft v6.21.

We first sought to identify and measure the strength of emission lines from the \SgrA\ accretion flow. To accomplish this goal, we chose the simplest model possible -- a power law -- to describe the continuum. We found that the zeroth order spectrum for \SgrA\ was well fit with a power law plus five gaussian emission lines. 
Table~\ref{tab:para_fit} shows the continuum fit parameters, line centers, and equivalent widths with the emitting elements.  The model fits with a reduced Cash statistic of 1.0. The measured line positions and equivalent widths are consistent with those reported in W13. The error bars on the iron line centroids and equivalent widths are likely smaller compared to W13 because this fit did not include an attempt to fit less prominent iron line features -- those of FeXXVI at 6.97~keV and neutral iron fluorescence at 6.4~keV.

\begin{table}
\centering
\caption{Parametric fit to quiescent \SgrA\ CCD spectrum}
\label{tab:para_fit}
\begin{tabular}{l c c c l}
  \hline
  \multicolumn{5}{c}{{\bf Model Parameter} (90\% CI)} \\
  \hline
  & ${\rm N}_{\rm H}$ & 10.69 & (10.08, 11.20) & $10^{22}$~cm$^{-2}$ \\
  & $\Gamma$          & 2.60  & (2.47, 2.70)   & \\
  \hline
  & \multicolumn{2}{c}{Line center [keV]} & \multicolumn{2}{c}{Equivalent Width [eV]} \\
  \hline
  Fe & 6.663 & (6.655, 6.672) & 726 & (659, 796) \\
  Ca & 3.890 & (3.866, 3.913) & 63  & (45, 88) \\
  Ar & 3.360 & (3.340, 3.380) & 77  & (55, 100) \\
  Ar & 3.074 & (3.050, 3.111) & 112 & (72, 163) \\
  S  & 2.429 & (2.400, 2.456) & 248 & (171, 342) \\
  \hline
\end{tabular}

\end{table}

W13 showed that the low resolution CCD spectrum of \SgrA\ can be fit with the RIAF model described in Equations~\ref{eq:ne} and \ref{eq:Te} with $q = 1.9 \pm 0.5$, implying $s \sim 1$.  Their spectral fits showed that the plasma temperature around the outer edge (Bondi radius) of the accretion flow was $kT \leq 1.3$~keV with a metal abundance about 1.5 times Solar.  
The best fit RIAF model from W13 also fits the CCD spectrum we extracted for \SgrA\ quiescent emission, with a reduced cash statistic of 1.1.

\subsection{Extraction of the \SgrA\ HETG Spectra}
\label{sec:HETGreduction}

The high energy transmission grating (HETG) instrument on \Chandra\ \citep{CanizaresHETG} obtains high resolution spectra by dispersing the X-ray light across an array of six CCD chips (ACIS-S). The HETG data processing pipeline employs an order-sorting algorithm that cross-references the energy of the photon event estimated from the amount of charge deposited in the CCD ($\sim 100$~eV accuracy) with the angular distance from the target of interest ($\sim 10$~m\AA when converting to wavelength).  The result -- in a vast majority of cases -- is a supremely low and therefore negligible background component. However, sources within $10''$ of \SgrA\ and the surrounding X-ray emitting plasma are similarly bright or brighter than \SgrA\ in quiescence. 

We used the CIAO tools \texttt{tg\_mask}, \texttt{tg\_resolve\_events} for order sorting, and \texttt{tgextract} for producing spectrum files for \SgrA.  To provide an accurate basis for comparison, we restricted the size of the source region in \texttt{tgextract} to a $1.5''$ cross-dispersion radius and the background regions to $1.5-5''$ in cross-dispersion angle. The background regions are extracted on both sides -- `up' and `down' -- of the dispersion direction axis.  

\begin{figure*}
	\includegraphics[width=\textwidth]{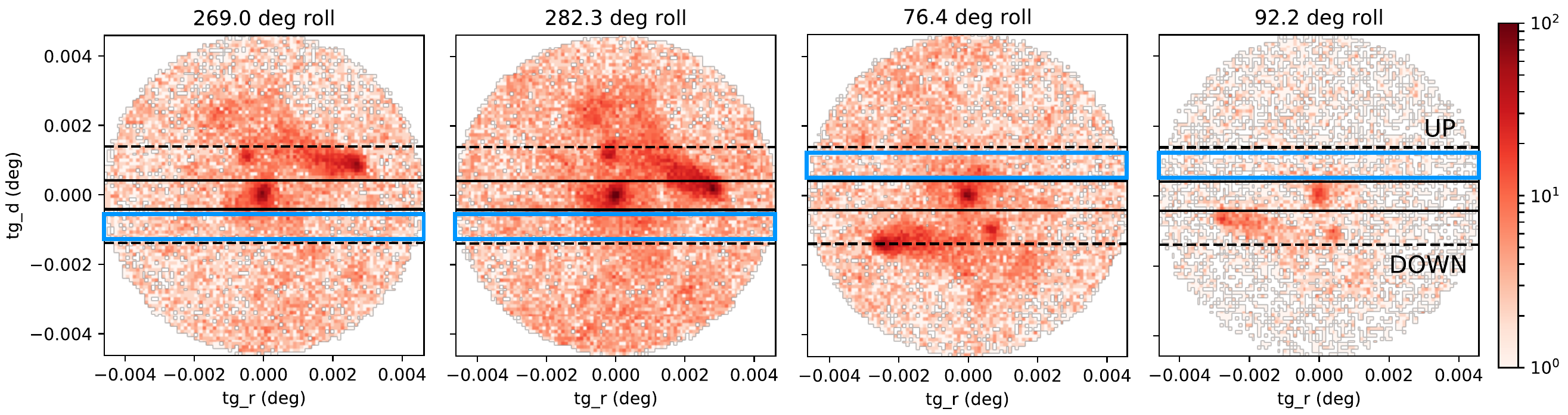}
	\caption{HETG zeroth order (non-dispersed) events histogram of \SgrA\ region stacked by roll angle, and displayed in HEG coordinates.  The horizontal axis (\texttt{tg\_r}) gives the angular distance from \SgrA\ in the HEG dispersion direction coordinates, and the vertical axis (\texttt{tg\_d}) corresponds to the cross-dispersion direction.  The $3''$ wide source extraction region is bracketed by the solid horizontal lines.  The dashed horizontal lines give the edges of the $1.5 - 5''$ region used to extract background. The GC X-ray emission sources \GCPWN\ and IRS~13 contaminate the `up' (positive \texttt{tg\_d}) background region for roll angles of 269 and 282 degrees, and the `down' (negative \texttt{tg\_d}) background region for roll angles of 76 and 92 degrees.  We use the opposite background regions, highlighted in blue, for modeling the background component to the HETG spectrum of \SgrA.
	}
	\label{fig:ZOimage}
\end{figure*}

The zeroth order image of two sources close to \SgrA, the pulsar wind nebular \GCPWN\ and likely star cluster IRS~13, lie within the background extraction region. The overlap is determined by roll angle (listed in Table~\ref{tab:hetgs_obsids}).  The GCXVP dataset employs four main roll angles, chosen so that the dispersion axis for \SgrA\ does not overlap with \SgrAEast, a supernova remnant with strong iron line emission.  We sorted the XVP observations into four roll angle categories for treating the background.  Figure~\ref{fig:ZOimage} shows the stacked zeroth order image of \SgrA\ after order sorting with \texttt{tg\_resolve\_events}, where \texttt{tg\_r} shows the angular position along the HEG arm dispersion direction, and \texttt{tg\_d} shows the angular position in HEG cross-dispersion coordinates.  Both \GCPWN\ and IRS~13 contaminate either the `up' and `down' background regions, but never both, allowing us to select the opposite as an appropriate background region for \SgrA.  

\subsubsection{Removing contaminating point sources}

The background component arises from several sources: (i) instrumental and particle background; (ii) dispersed light from sources near \SgrA\ in angular distance; (iii) the zeroth (non-dispersed) image of the entire GC that lies within the order sorting window; and (iv) HETG dispersed light from GC sources that overlap with the dispersed light from \SgrA.

Subtracting the HETG extracted spectrum from an adjacent region (background `up' or `down') will remove nearly all the effects {\sl except} in cases where a point source overlaps with the source region and not the background region, or vice versa. The result is an emission (or absorption) artifact that might be interpreted as a line feature. For example, we observed a 3~\AA\ artifact feature in the MEG-1 spectrum that arose from an X-ray transient in the \SgrA\ extraction region, for the epoch with roll angle of 76~degrees. Its angular distance from \SgrA, $2.6'$, corresponds to the dispersion angle for 3~\AA\ photons.

We separated the observations into four epochs, based on roll angle, and ran CIAO \texttt{wavdetect} on the stacked images to identify X-ray sources on wavelet scales of $0.5'' - 2''$. We wrote a Python script to transform the \texttt{wavdetect} output into gratings coordinates for MEG and HEG, and flagged sources that overlapped with the source extraction region or the background regions for the respective epoch (described above). For the 2-8~keV range, the furthest contaminating point source has an off-axis angle of approximately $600''$ (HEG) or $300''$ (MEG), for which the corresponding 90\% encircled energy fraction (EEF) for the 1.5~keV PSF is approximately $11.5''$ and $4''$, respectively. We used these angular values to construct ignore regions for wavelengths within the 90\% EEF of each point source that overlapped with the HETG spectral extraction regions.

We did not calculate ignore regions for case (iv), where the MEG (or HEG) dispersed spectrum from a different source overlaps with the \SgrA\ HEG (or MEG) arm. First, the HEG and MEG effective area is 3-5 times lower than the zeroth order effective area, making the effect small in comparison to removing zeroth order point sources. 
Second, there was no point source in the dataset that was significantly bright enough for the HETG dispersed spectrum to outshine the zeroth order image of diffuse GC emission plus particle background. Overall, we estimate that the effect of HETG dispersed light from other sources overlapping with the HETG dispersed spectrum of \SgrA\ is negligible compared to the other three sources of background.

\subsubsection{Stacking the spectra}

We wrote an S-lang script to stack the source and background spectra with ISIS. We made a combined counts histogram by stacking the raw counts from every observation. We assume that the HETG RMF -- which maps dispersion angle to photon energy while accounting for the grating line-spread-function -- is relatively stable, and adopt the ObsID 14439 RMF for the combined dataset. We used an exposure weighted average to stack the HETG ARF -- which gives the effective instrument area as a function of energy -- into a combined ARF. For the combined background, we stacked the `down' background from datasets with roll angles 269 and 282 degrees with the stacked `up' background from datasets with roll angles 76 and 92 degrees.  Counts and ARF values in the ignored wavelength regions were set to zero before stacking. 
Figures~\ref{fig:HEGfitspectra} and \ref{fig:MEGfitspectra} show the scaled background datasets (dark blue) overlaid with the raw \SgrA\ spectrum (black) for the HEG and MEG datasets, respectively.

\section{Results}
\label{sec:Results}

We analyzed the stacked spectrum and response files using a custom Python library\footnote{github.com/eblur/pyxsis} in order to take advantage of publicly available Python modules for scientific analysis: Scipy \citep{Scipy}, Astropy \citep{Astropy2}, and Emcee \citep{emcee}. 
Due to the low number of counts in each bin, we plot the HEG and MEG data binned by a factor of 20 and 10, respectively. However, we evaluate the \SgrA\ fit statistics on unbinned data, as described below.

When it comes to fitting the background, we sought a reliable approximation for the background spectral shape, which cannot be described by an analytic or physical model. 
We modeled each background dataset by applying a cubic spline to the binned counts histogram. MEG and HEG background datasets were binned by a factor of 10 and 20, respectively. We used the Scipy UnivariateSpline function with a smoothing factor of 100 and used the Gehrels error as a weighting factor. The spline was then interpolated to create a background model for the unbinned counts histogram.

All fits for the \SgrA\ spectrum were evaluated with the Cash statistic on the unbinned counts histogram, after adding the background model. We utilized Bayesian fitting techniques by calculating a posterior distribution for the model likelihood using the MCMC ensemble sampling code, \textit{emcee} \citep{emcee}.
We fit the +1 and -1 orders of each dataset separately in order to prevent instrumental artifacts or Poisson fluctuations to be mistakingly attributed to absorption or emission lines. To be considered real, a residual feature must appear in both spectra.

\subsection{Fit to HEG Spectra with Iron Line}
\label{sec:HEGFit}

\begin{figure*}
\centering
	\includegraphics[width=0.48\textwidth]{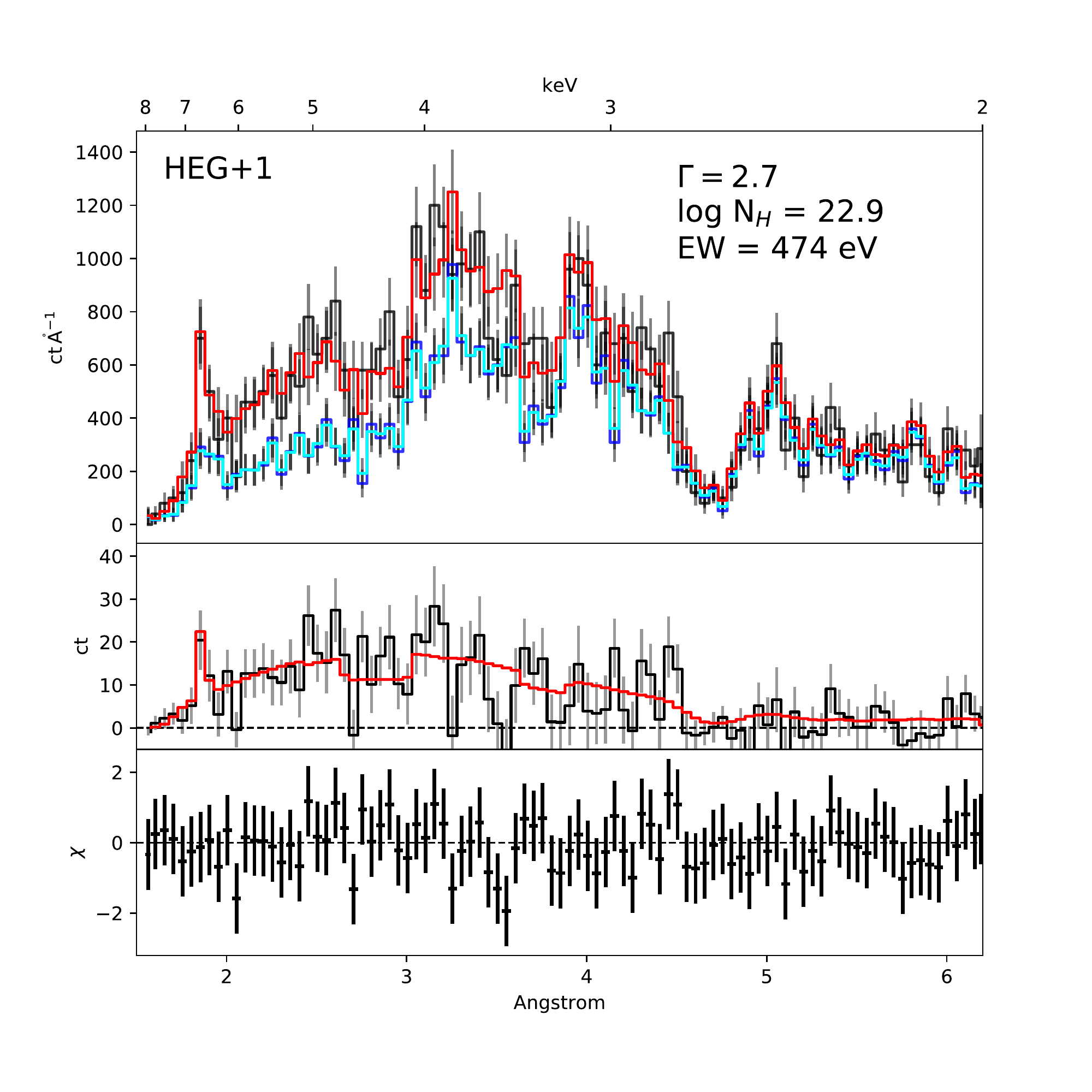}
	\includegraphics[width=0.48\textwidth]{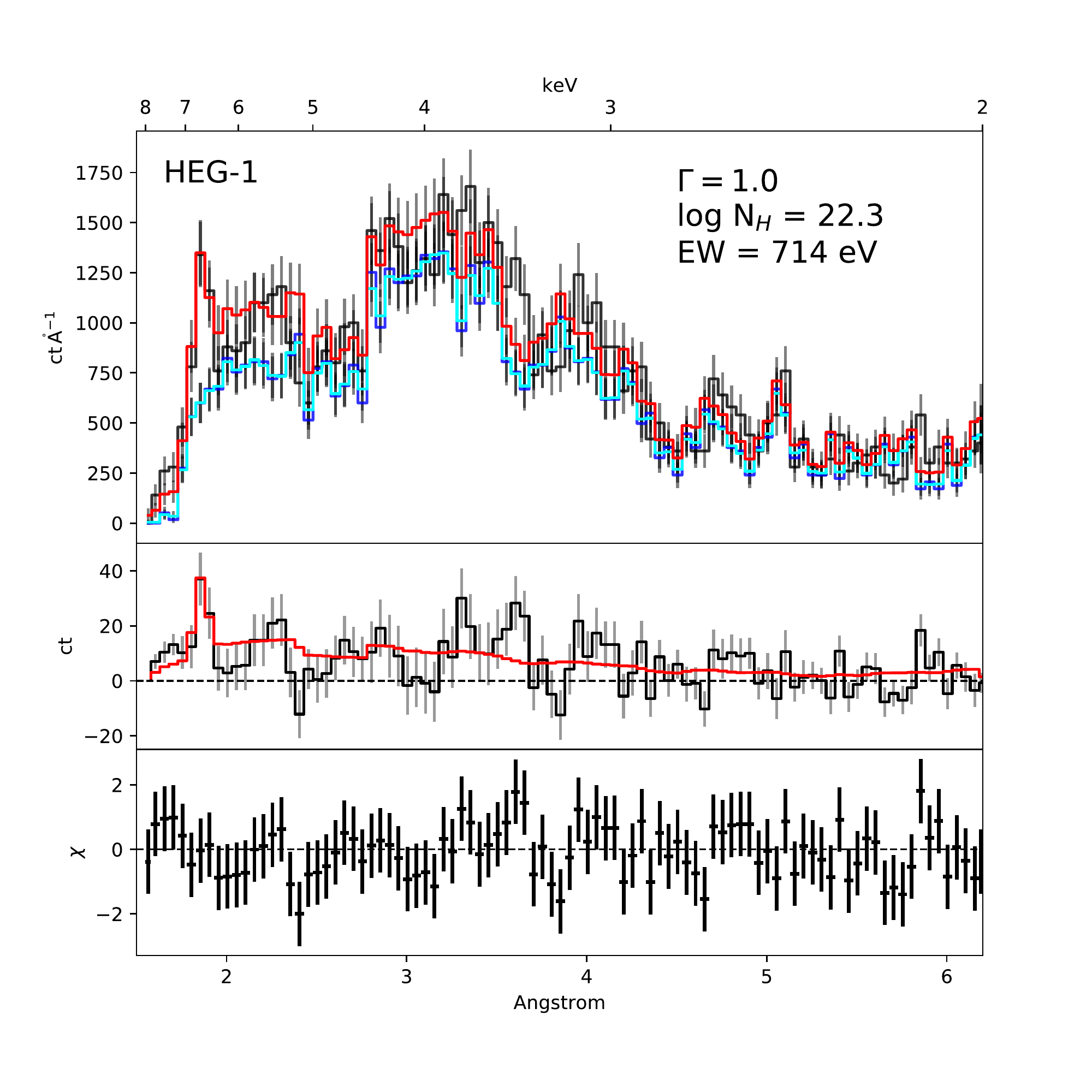}
	\caption{Best fit {\sl emcee} walker for the HEG dataset. 
		\textit{Top panels:} The raw \SgrA\ counts histogram is plotted in black and the raw background is plotted in dark blue, and both are binned by a factor of 20. The background spline model is overlaid in light blue, and the background plus model is overlaid in red. 
		\textit{Middle panels:} The background subtracted, binned \SgrA\ counts histogram (black) is plotted for visual inspection of the binned model counts (red). 
		\textit{Bottom panels:} Model residuals divided by the Gehrels error bar are plotted. There are no significant emission or absorption line features that appear in both spectra.
		}
	\label{fig:HEGfitspectra}
\end{figure*}

We modeled \SgrA\ with a power law plus a Gaussian iron line under the effects of ISM extinction, using the same models described in Section~\ref{sec:CCDspectrum}. 
We placed several Gaussian priors on model parameters and outcomes. 
The foreground ISM column received a Gaussian prior of ${\rm N}_{\rm H} = (15 \pm 2) \times 10^{22}$~cm$^{-2}$.  The  2-10~keV luminosity calculated from each set of parameters also received a Gaussian prior of $(2.4 \pm 1.0) \times 10^{33}$~erg~cm$^{-2}$~s$^{-1}$. 
Table~\ref{tab:PosteriorHEG} summarizes the final results for the HEG+1 and HEG-1 model posteriors, with 90\% confidence intervals in parenthesis. The photon index and ISM column measurements differ significantly between the two spectra, but the iron line centroid and equivalent widths agree completely. We consider the HEG+1 spectrum a more reliable measurement of the \SgrA\ continuum, because it agrees better with fits made to the CCD spectrum (Table~\ref{tab:para_fit}) and the ISM column is in greater agreement with values reported in the literature for GC objects \citep[e.g., W13;][]{Baganoff2003,Nowak2012,Ponti2016}. 
Figure~\ref{fig:HEGfitspectra} show the spectra and residuals using the best fit walkers for the HEG+1 and HEG-1 spectra. There are no residual emission or absorption features that appear concurrently in both spectra.

\begin{table*}
\centering
\caption{Posterior distribution of HEG model parameters with 90\% confidence intervals}
\label{tab:PosteriorHEG}
\begin{tabular}{c l l l l l l l l l l} 
\hline 
	 	 &	 \multicolumn{2}{c}{${\rm N}_{\rm H}$}  &	 \multicolumn{2}{c}{{Photon Index}}  &	 \multicolumn{2}{c}{{Fe centroid}}  &	 \multicolumn{2}{c}{{Fe line EW}}  &	 \multicolumn{2}{c}{{$L_{X} (2-10~{\rm keV})$}}  \\ 
	 	 &	 \multicolumn{2}{c}{$10^{22}$ cm$^{-2}$}  &	 \multicolumn{2}{c}{$\Gamma$}  &	 \multicolumn{2}{c}{keV}  &	 \multicolumn{2}{c}{eV}  &	 \multicolumn{2}{c}{$10^{33}$ erg cm$^{-2}$ s$^{-1}$}  \\ 
\hline 
	 HEG+1  &	 8.4  &  (5.9, 10.8)  &	 2.6  &  (2.1, 3.2)  &	 6.652  &  (6.599, 6.695)  &	 360  &  (131, 656)  &	 3.4  &  (2.7, 4.3)  \\ 
	 HEG-1  &	 2.3  &  (1.0, 3.8)  &	 1.1  &  (1.0, 1.5)  &	 6.681  &  (6.629, 6.730)  &	 658  &  (362, 1002)  &	 2.1  &  (1.8, 2.3)  \\ 
\hline 
\end{tabular}
\end{table*}

We quantified the significance of the iron line detection by performing a null hypothesis test. We fit the dataset with a null hypothesis (power law only), and calculated the likelihood ratio between the line model and the power law without a line. We then simulated 500 fake datasets using the HEG$\pm$1 response files using the best fit power law and no Gaussian line. The simulated dataset was fit with and without a 6.7~keV Gaussian feature, and we computed the likelihood ratio between the two models. Comparing this to the likelihood ratio measured from the dataset, there is $< 3\%$ probability that the null hypothesis is correct.

\subsection{Fit to MEG Spectra}
\label{sec:MEGFit}

The \Chandra\ MEG array does not have sufficient sensitivity in the 6-8~keV energy range to obtain significant results for the iron emission line from \SgrA. However, its moderate energy sensitivity could allow us to constrain other lines expected to be present (\S~\ref{sec:CCDspectrum}). We applied the same modeling techniques and priors as those used for the HEG spectra (\S~\ref{sec:HEGFit}), with the Gaussian line omitted. 

\begin{figure*}
\centering
	\includegraphics[width=0.48\textwidth]{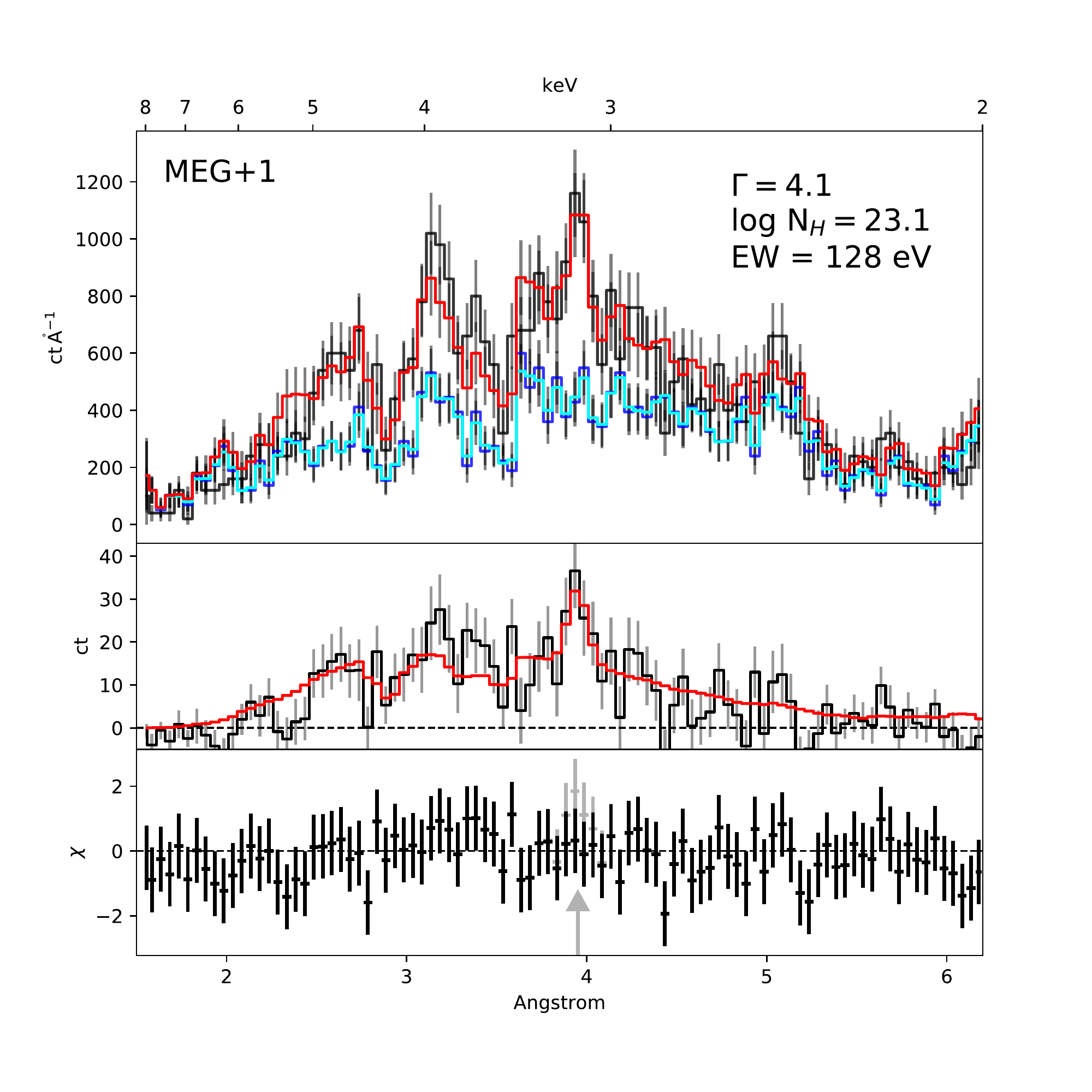}
	\includegraphics[width=0.48\textwidth]{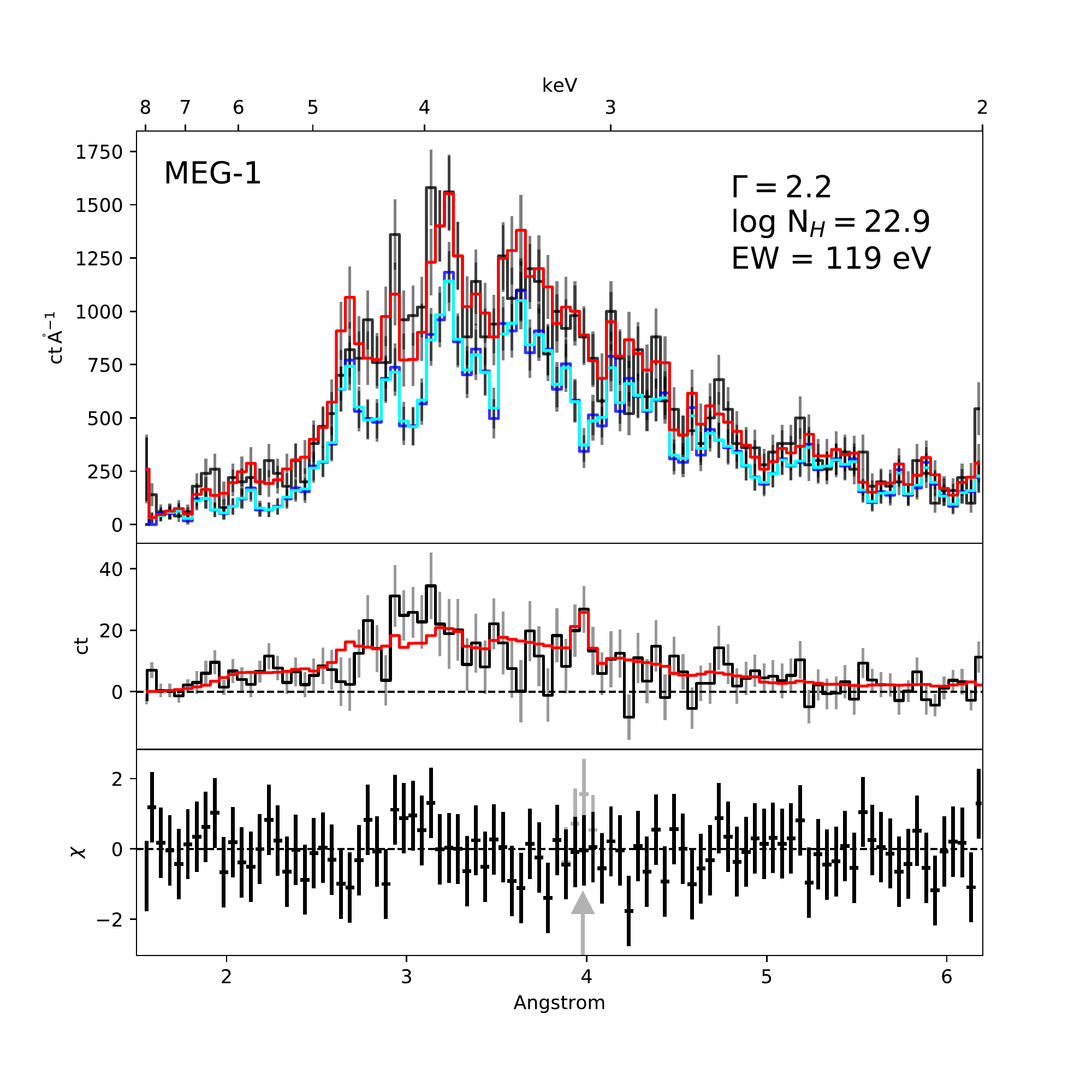}
	\caption{Best fit {\sl emcee} walker for the MEG dataset, plotted in the same way as Figure~\ref{fig:HEGfitspectra}. 
		The MEG models do not incorporate a Gaussian iron line features, because the MEG gratings do not have sufficient sensitivity. When the spectra are fit with a power law only, both sets of residuals (bottom panel, grey error bars) show indications of an emission feature around 3.95~\AA, likely associated with Argon, indicated with the grey arrow. When we freeze the power law to its best fit values, the parameters described in the inset text, and fit for a Gaussian emission lines, the residuals improve (bottom panel, black error bars). The centers of the fit lines are marked with a grey arrow.
		}
	\label{fig:MEGfitspectra}
\end{figure*}

The left hand portion of Table~\ref{tab:PosteriorMEG} summarizes the final results for the power law and ISM extinction models fits to the MEG data. The MEG+1 dataset was well constrained with a single population of model parameters, but the MEG-1 dataset exhibited a multiple local maxima to the posterior probability distribution. We chose the population of {\sl emcee} walkers that best agreed with previous luminosity measurements of \SgrA, i.e. $\log {\rm L}_X < 33.8$ in erg~cm$^{-2}$~s$^{-1}$.  Figure~\ref{fig:MEGfitspectra} show model spectra obtain from the best fit walkers. The grey residuals in the bottom panel of both figures hint at an emission line around 3.95~\AA (indicated with an arrow), likely associated with Argon.

With the power law parameters frozen to the best fit values, we re-ran {\sl emcee} to fit a Gaussian line. We limited the Gaussian center to the range of 3.6-4.1~\AA\ and applied the prior to the \SgrA\ luminosity as described for the power law fit, above. The resulting fits show improved residuals (shown in black, bottom portion of Figure~\ref{fig:MEGfitspectra}) with a line centered on 3.1~keV with an equivalent width of 60-180~eV (middle portion of Table~\ref{tab:PosteriorMEG}).

\begin{table*}
\centering
\caption{Posterior distribution of MEG model parameters with 90\% confidence intervals}
\label{tab:PosteriorMEG}
\begin{tabular}{c l l l l | l l l l | l l} 
\hline 
	 	 &	 \multicolumn{2}{c}{${\rm N}_{\rm H}$}  &	 \multicolumn{2}{c}{{Photon Index}}  &	 \multicolumn{2}{c}{{Line centroid}}  &	 \multicolumn{2}{c}{{Line EW}}  &	 \multicolumn{2}{c}{{$L_{X} (2-10~{\rm keV})$}}  \\ 
	 	 &	 \multicolumn{2}{c}{$10^{22}$ cm$^{-2}$}  &	 \multicolumn{2}{c}{$\Gamma$}  &	 \multicolumn{2}{c}{keV}  &	 \multicolumn{2}{c}{eV}  &	 \multicolumn{2}{c}{$10^{33}$ erg cm$^{-2}$ s$^{-1}$}  \\ 
\hline 
	 MEG+1  &	 12.7  &  (11.6, 13.8)  &	 4.1  &  (3.7, 4.4)  &	 3.140  &  (3.115, 3.166)  & 124  &  (74, 183)  &  6.2  &  (5.2, 7.3)  \\ 
	 MEG-1   &	 8.8    &  (8.0, 9.6)  	 &	 2.4  &  (2.2, 2.7)  &	  3.115  &  (3.097, 3.131)  &  111  &  (57, 170) &  4.1  &  (3.7, 4.6)  \\ 
\hline 
\end{tabular}
\end{table*}

We quantified the significance of the line detection by performing a null hypothesis test, following the techniques used for the HEG dataset. We simulated 500 fake datasets using the MEG$\pm$1 response files and the best fit power law with no Gaussian line. The simulated dataset was fit with and without a 3.1~keV Gaussian feature, and we computed the likelihood ratio between the two models. Comparing this to the likelihood ratio measured from the dataset, there is $< 10\%$ probability that the null hypothesis is correct.

\section{Discussion}
\label{sec:Discussion}

As noted above, the continuum model of a power law does not match well to the observed interstellar extinction towards other Galactic Center sources, which typically have $\NH = 1.5 \times 10^{23}$~cm$^{-2}$. Similar results are achieved when modeling the emission of \SgrA\ with a single temperature plasma (W13). Thus a more complex continuum model is needed to describe the accretion flow. Despite the range of parameters used to describe the underlying continuum, we are able to measure two emission lines: the 6.7~keV iron line in both the HEG$\pm$1 spectra, and a 3.1~keV line (likely argon) in both of the MEG$\pm$1 spectra. 

\subsection{Comparison of Gaussian line fits to theoretical predictions}
\label{sec:GaussCompare}

\begin{figure}
\centering
	\includegraphics[width=0.48\textwidth]{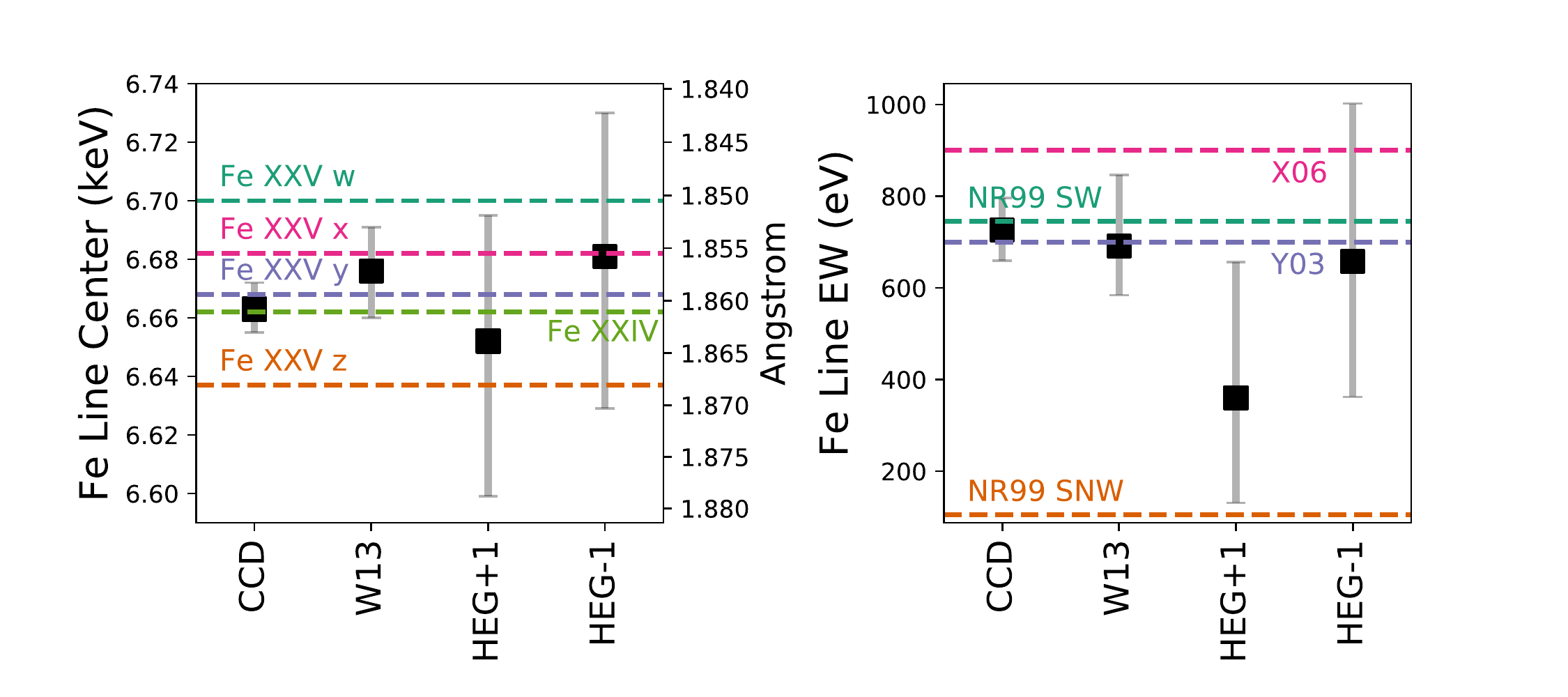}
	\caption{
		Measurements of the iron line centroid (left) and equivalent width (right) from the parametric fit to the CCD spectrum in this work (Table~\ref{tab:para_fit}), the W13 parametric fit to the CCD spectrum, and the HEG$\pm$1 orders (Table~\ref{tab:PosteriorHEG}).
		(Left:) The position of the He-like iron triplet (FeXXV w,x,y,z) lines, which have the largest emissivity values in the 6.6-6.8~keV range (AtomDB 3.0.9), are marked by the horizontal dashed lines. An FeXXIV line, within the region of the FeXXV triplet, is also marked, but has a lower emissivity. 
		(Right:) The horizontal dashed lines mark the equivalent widths for FeXXV predicted by several theoretical works. X06 provided predictions for the FeXXV equivalent width using the Y03 predicted value of $s=0.3$ (line marked Y03), as well as $q/s \sim 1$ (line marked X06). The X06 model shown here provided the smallest equivalent width prediction, after Y03. The NR99 work utilized an ADAF model with outflows (SW) and without outflows (NSW), and predicted equivalent widths for each line of the FeXXV triplet. The horizontal lines show the sum of equivalent widths in the triplet, showing that accretion flow models with outflows are more consistent with the observed Fe line equivalent width than those without. 
		}
	\label{fig:FeLine}
\end{figure}

Figure~\ref{fig:FeLine} visualizes the results for the iron line emission described in this paper and W13. We overlay all of the strong lines (with emissivity $> 10^{-18}$~phot~cm$^3$~s$^{-1}$) listed in AtomDB~3.0.9. This is the most up-to-date version of AtomDB, for which the position of He-like iron lines have been adjusted as a result of the high-resolution X-ray spectra obtained from {\sl Hitomi} observations of, e.g., the Perseus cluster \citep{HitomiPerseusAtomDB}. The measurements of the HEG$\pm$1 datasets are consistent with a blend of the FeXXV triplet, for which the resonant (w) and forbidden (z) lines are strongest. While it is in this range of energies, predictions for the the FeXXIV line equivalent width 
are not described in any of the theory work evaluated below.

The right side of Figure~\ref{fig:FeLine} shows the measured FeXXV equivalent width and comparisons to various predictions for the \SgrA\ accretion flow. 
\citet[][hereafter NR99]{Narayan1999} used an ADAF model ($q \approx 1$) and altered the mass accretion profile to account for the presence of a wind (SW, yielding $s \approx 1/2$) or no wind (SNW, yielding $s \approx -1/2$). The wind models lead to lower temperatures throughout a larger volume of the accretion flow, producing to higher equivalent widths overall. Figure~\ref{fig:FeLine} shows the NR99 predicted sum of equivalent widths from all lines in the He-like iron doublet, from models with and without winds. As expected, the no-wind case of NR99, which has higher temperatures overall, is ruled out. 

\citet[][hereafter Y03]{Yuan2003} developed a RIAF model 
to fit the multi-wavelength quiescent spectrum of \SgrA\ along with the constraints made on the electron density from radio observations. 
Their radiative transfer model includes both thermal and non-thermal populations of electrons, the latter of which are accelerated through magnetic turbulence. The best fitting model had a temperature and density distribution that follows $q \approx 3/4$ and $s \approx 1/2$, respectively. 
\citet[][hereafter X06]{Xu2006} provided predictions for the thermal X-ray line emission from the generalized RIAF model described in \S~\ref{sec:Intro} and the Y03 model specifically. The magenta dashed line shows the X06 results for $q = 0.75$ and $s = 0.75$, which is the only X06 model that agrees with both the results shown in Figure~\ref{fig:FeLine} and the radio measurements for free electron density in the accretion flow.

The iron line centroids and equivalent widths obtained from the HEG measurements are not as well constrained as the fits obtained with the CCD spectrum. As we will show in \S\ref{sec:RIAFCompare}, this spread likely due to the complex of iron lines that appear with low signal-to-noise, making fits with a single Gaussian line inappropriately broad. Considering that past theoretical works did not provide comprehensive predictions for a variety of iron ions, we reserve this evaluation for a later work.

\begin{figure}
\centering
	\includegraphics[width=0.48\textwidth]{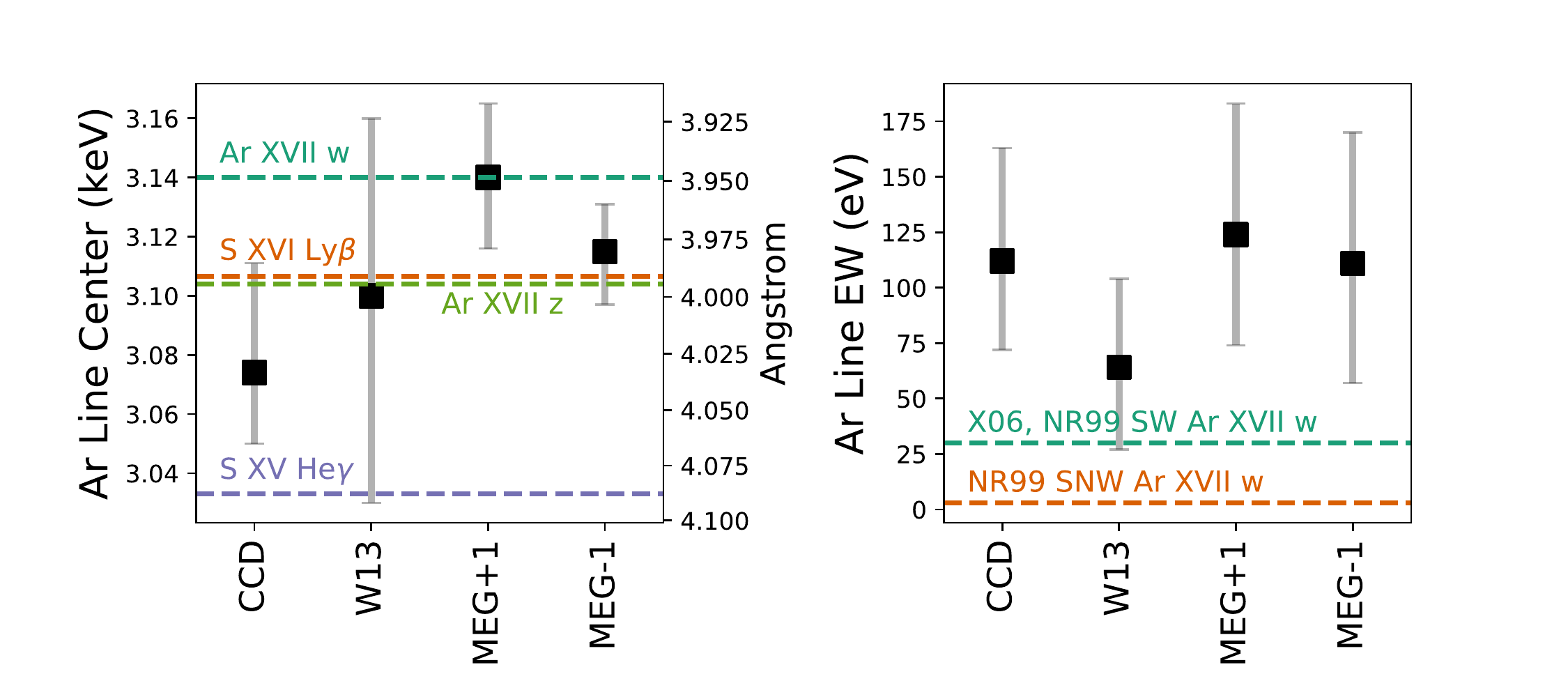}
	\caption{
		Measurements of the Ar line centroid (left) and equivalent width (right) from the parametric fit to the CCD spectrum in this work (Table~\ref{tab:para_fit}), the W13 parametric fit to the CCD spectrum, and the MEG$\pm$1 orders (Table~\ref{tab:PosteriorMEG}). 
		(Left:) The position of five lines in the 3.0-3.2~keV energy range with the largest emissivity values, as reported by AtomDB 3.0.9. The highest emissivity comes from ArXVII, which peaks in abundance for temperatures around 1.7~keV. The 3p to 1s transition from H-like sulfur (SXVI Ly$\beta$) has a similar emissivity to the Ar doublet, but peaks for temperatures around 2.2~keV. 
		(Right:) The horizontal dashed lines mark the equivalent widths for ArXVII modeled by NR99 and X06. The measured equivalent widths all exceed the predictions from the \SgrA\ models with outflows (NR99 SW and X06, which provide the same prediction) and without winds (NR99 SNW). This is an indication of super-Solar abundance of argon in the accretion flow of \SgrA.
		}
	\label{fig:ArLine}
\end{figure}

Figure~\ref{fig:ArLine} visualizes the fit results for the 3.1~keV emission line, which is also measured in W13. The MEG dataset provides a stronger constraint on the line center than that achieved by W13, while obtaining similar results for the equivalent width. The strongest emission lines, as listed in AtomDB~3.0.9, between 3.0 and 3.2~keV are overlaid with the measurements for the line center in Figure~\ref{fig:ArLine}, left. The He-like Ar triplet aligns well with the MEG measurements. H-like sulfur (S~XVI) has emissivity values that are $\sim 30\%$ that of the the Ar XVII triplet. The ion abundance for S~XVI peaks at a similar temperature (2.2~keV) to Ar~XVII (1.8~keV), leading to the possibility that both are present.
However, due to the relative emissivity values, we presume that argon dominates; we follow the precedent set by NR99, X06, and W13 by referring to this line as arising from argon.

The right side of Figure~\ref{fig:ArLine} shows the equivalent width measurement for the Ar line along with predictions made by X06 and NR99. All measurements of the Ar line EW are consistent with each other, and generally higher than  predicted. 
All of the above theoretical works utilized Solar abundances. The RIAF model presented by W13 fit best with metal abundances between 1 and 2 times Solar. Assuming a Solar mix of abundances, supersolar abundance in the accretion flow of \SgrA\ would explain the enhanced emission from argon but not that of iron. This would imply that iron is depleted relative to other metals in the GC region, such as observed in the accretion material of AX~J1745.6-2901 \citep{Ponti2016}.

\subsection{Comparison to RIAF models of W13}
\label{sec:RIAFCompare}

We stacked the HEG$\pm 1$ spectra and forward modeled the best fit W13 RIAF to predict the corresponding counts histogram. 
In this model, the density profile is relatively flat ($s \sim 1$) and $q \geq 0.6$. The continuum is dominated by bremsstrahlung emission from hot inner layers, and the line emission is dominated by cooler outer regions of the accretion flow. The W13 model fits with a reduced Cash statistic of 3.2; in contrast, the power law plus Gaussian line models have a reduced Cash statistic around 2.5.

\begin{figure}
\centering
	\includegraphics[width=0.48\textwidth]{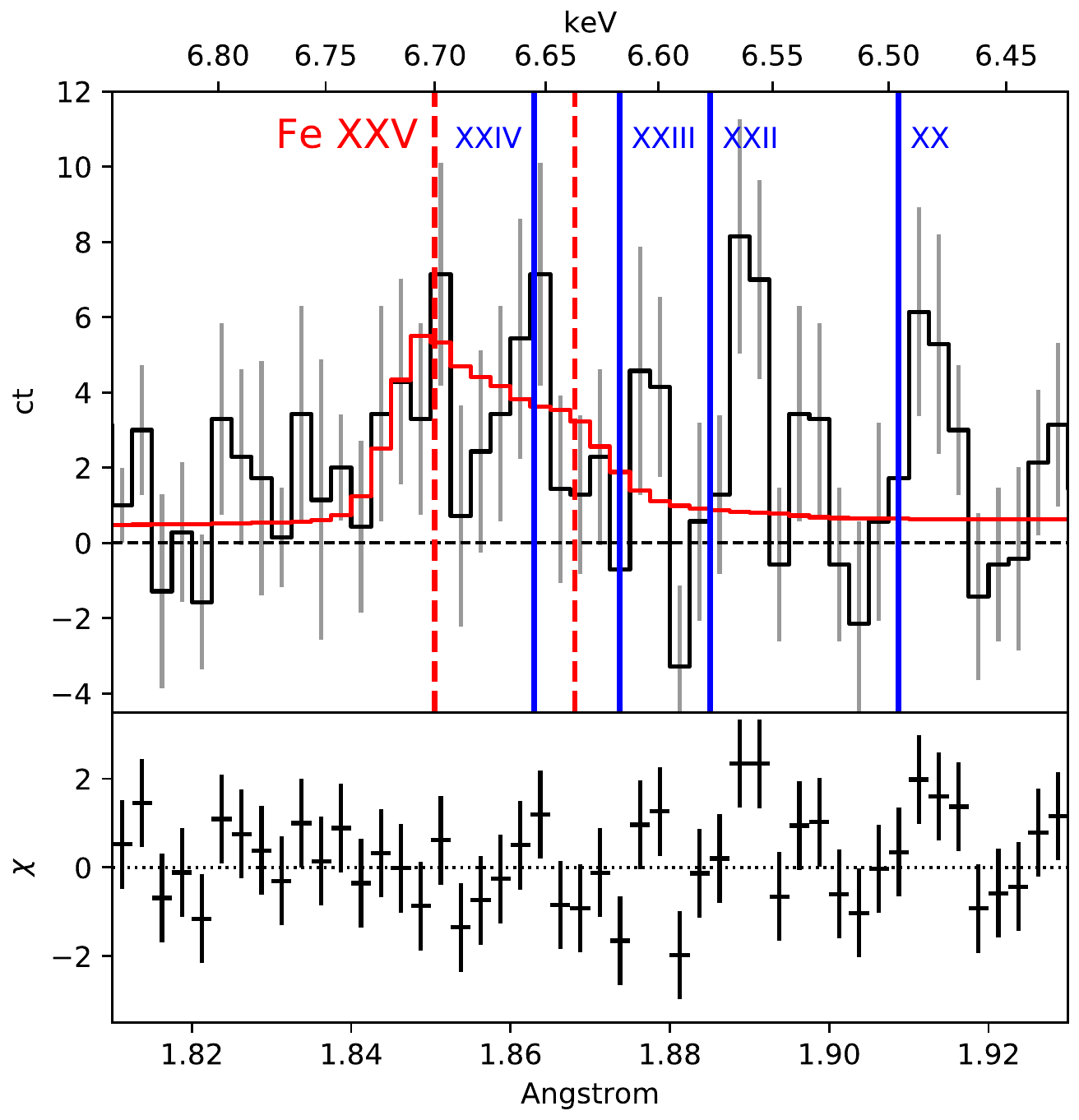}
	\caption{
		Iron line region of unbinned HEG dataset, $\pm 1$ orders stacked and background subtracted. The best fit RIAF model from W13 is overlaid in red, and the residuals are shown in the bottom portion of the plot. The FeXXV doublet is present in our dataset (red dashed lines), but there are additional features that appear with higher signal-to-noise than the He-like iron emission features. It's possible that these lines arise from lower ionization states of iron (blue lines), associated with a cooler ($\sim 1$~keV) plasma.
		}
	\label{fig:W13RIAF}
\end{figure}

Figure~\ref{fig:W13RIAF} shows the unbinned HEG$\pm 1$ spectrum, background subtracted, in the Iron emission line region. The W13 model, shown in red, shows two distinct peaks, which correspond to the resonant and forbidden lines of the He-like iron triplet (FeXXV~w and z in Figure~\ref{fig:FeLine}, left). The stacked spectrum also shows two peaks that roughly correspond to these two lines. However, higher signal-to-noise peaks are visible at longer wavelengths than expected from FeXXV emission. A single temperature APEC model fit to the 1.5-2~\AA\ spectrum suggests that these lines could arise from lower ionization states, FeXX$-$FeXIV (blue lines), arising from a 0.9~keV plasma. 
In agreement with W13, we find no indication of neutral Iron fluorescence (FeI K$\alpha$ doublet at 1.936 and 1.940~\AA). 
We cannot provide constraints on the presence of the 7.881~keV Fe~XXV line, detected by W13, because the effective area of the HEG at that energy is too low to detect more than 1-2 photons from that line.

The lower temperature iron lines likely arise from material that is further away from the center of \SgrA. In the cross-dispersion direction, the HETG spectra are limited to gas within $1.5''$ of \SgrA, the size of the region chosen in W13. With a black hole mass of $4.02 \times 10^{6}~M_{\odot}$ and distance of $7.86$~kpc \citep{Boehle2016}, $1.5''$ corresponds to $1.5 \times 10^5$ Schwarzschild radii ($r_s$) in projection. However, due to the fact that the HETG is a slit-less spectrometer, more image area in the dispersion direction might be contributing to the high resolution spectrum. 

To evaluate this possibility, we calculated the dispersion angles associated with the energy window used by the HETG order-sorting algorithm for the photon energy of interest. Assuming a 200~eV order-sorting window at 6.7~keV, the HEG spectrum samples the zeroth order image from $\pm 3.3''$ in the dispersion direction.  This corresponds to $r < 3 \times 10^5 r_s$ and is slightly beyond the Bondi radius, which is $3''$, assuming that the sound speed of the ambient material is $\approx 550$~km/s \citep{Baganoff2003}. Overall, the detection of emission from lower ion states of iron are in agreement with W13, which found evidence that the iron emission lines are shifted towards lower energy in the spectrum sampled from a $2-4''$ annulus around \SgrA.

The unbinned HEG$\pm 1$ spectrum exhibits much structure in the iron line region that requires multi-temperature plasma modeling along with special consideration of the spatial regions being sampled, described below. 
Note that binning the HETG spectra to achieve a higher signal-to-noise counts histogram smooths out the iron line structures, as seen in Figure~\ref{fig:HEGfitspectra}, limiting the diagnostic potential. Additionally, and more importantly, many of the counts histogram peaks are not centered on the predicted iron line centroids. This is an indication that velocity structure must be considered in modeling the accretion flow of \SgrA. Complicating this matter is the fact that the bulk of theoretical line centroids and emissivities provided in atomic databases, especially for moderate ionization states, have yet to be calibrated and tested in the lab. 
It has been demonstrated that high resolution spectra of astronomical objects can be used to calibrate these values \citep{Garcia2005, HitomiAtomDB}, providing the opportunity for the spectrum of \SgrA\ to contribute to fundamental physics. 
We reserve a deeper look at the \SgrA\ iron line complex and a search for additional lines for a subsequent paper.

Some systematic effects influence the appearance of Figure~\ref{fig:W13RIAF}. Because the background extraction regions are adjacent to the source extraction region (Figure~\ref{fig:ZOimage}), 
lines from the $1.5-3.3''$ accretion flow captured in the HETG dispersion direction are affected by subtraction of the background region, which contains lines from the $1.5-5''$ accretion flow captured in the HETG cross-dispersion direction.  Thus explaining the apparent shifts in the Fe~XX-XIII lines will require a model that takes into account the orientation of the rotational velocity structure relative to the HETG dispersion axes. 
Future work can quantify this effect by examining background regions that are further away in the cross-dispersion direction, or by fitting the background and source spectra simultaneously. It is also important to note that, by choosing a background region that is always to the east of \SgrA, these results are subject to bias from potential asymmetries in the GC gas distribution.

\section{Conclusions}
\label{sec:Conclusions}

With this work, we are publicly releasing the stacked foreground and background HETG spectra for \SgrA\ quiescent emission obtained via the {\sl Chandra} GCXVP campaign.\footnote{Available on the {\sl Chandra} contributed datasets website: http://cxc.cfa.harvard.edu/cda/Contrib/2020/CORR1/} \SgrA\ flares detected in \citet{Neilsen2013} have been removed, as well as point sources in the zeroth order image that contaminate the dispersed HETG spectra. The final dataset offers a 2.5~Ms exposure of \SgrA's 2-8~keV X-ray emission with $R \sim 500$ spectral resolution.

After applying a cubic spline to interpolate the background counts histogram, we fit a parametric model of a power law plus Gaussian lines using the unbinned spectra. We detect 6.7~keV (1.85~\AA) iron emission in the HEG dataset and a 3.1~keV (3.96~\AA) emission line in the MEG dataset, likely dominated by argon. The line centroids and equivalent widths are consistent with measurements obtained from the low resolution CCD spectrum and with the analysis by W13. In the case of iron, a simple parametric fit to the HEG data provides no better constraints. In the case of the argon line, we obtain a better constraint on the line centroid, confirming that it is most likely arising from the He-like argon doublet. Overall, both the iron and argon line measurements agree more readily with RIAF models where $s \sim 1/2 - 1$, caused by outflow.

The stacked HEG$\pm 1$ shows substantial iron line structure arising from multi-temperature gas layers, sampling the accretion flow out to the Bondi radius ($r < 3 \times 10^5 r_s$). Predictions for these line strengths have not been made in previous works, which focus mainly on the accretion flow within $10^5 r_s$ ($<1-1.5''$). Modeling the temperature structure for this dataset will have to take into account the portion of image sampled as a result of the {\sl Chandra} HETG order sorting algorithm, which compares dispersion angle with the CCD pulse height and HETG line spread function in order to assign photon events to a higher resolution energy bin.

The fact that the iron lines arising from lower ionization states, FeXX-XXIV, are not centered on the AtomDB values suggests the possibility of complex velocity structure. This possibility is supported recent results of \citet{Ma2019}, which modeled the multi-wavelength SED for the inner accretion flow ($r < 10^3 r_s$) of \SgrA. They show evidence that the density structure for the central portion of the accretion flow ($s < 1$ for $r < 10^3 r_s$), which accounts for 4\% of the total quiescent X-ray emission, is different from that seen for outer layers of the flow ($s \sim 1$ for $10^3 - 10^5 r_s$). These results provide support for theoretical models that find different outflow velocities, and thus a changing $s$ value, depending on the distance from \SgrA\ \citep[e.g.,][and references therein]{Yuan2012a,Yuan2015}. 
We reserve an examination of individual line features in the unbinned HEG and MEG spectra, their kinematics, and more extensive RIAF modeling for a followup paper.

\acknowledgements
We thank all members of the \SgrA\ {\Chandra}-XVP team. We would also like to thank Daniela Huppenkothen (Dirac Institute at the University of Washington) and Roman Garnett (Washington University in St. Louis) for useful conversations about statistics and Python coding for X-ray spectroscopy. 
We thank the anonymous referee for their helpful comments and clarifying questions, which strengthened the published results. 
This research was made possible by the {\sl Chandra} X-ray Visionary Program supported by the Chandra X-ray Center (CXC), which is operated by the Smithsonian Astrophysical Observatory for NASA under contract NAS8-03060. Additional support for this work was provided by NASA through Einstein Postdoctoral Fellowship grant number PF6-170149, awarded by the CXC. 
All figures in this work were created with Python library Matplotlib \citep{Matplotlib}.

\bibliography{/Users/lia/Dropbox/Bibliographies/references_new}

\end{document}